\documentclass[twocolumn, showpacs,amsmath,amssymb,aps,prl,floatfix, 10pt]{revtex4-1}
\usepackage[latin1]{inputenc}
\usepackage[american]{babel}
\usepackage[T1]{fontenc}
\usepackage{mathrsfs}
\usepackage{hyperref}
\usepackage{multirow}
\usepackage{bm, epsfig}
\usepackage{graphicx}
\usepackage{subeqnarray}
\usepackage{bbold}
\usepackage{upgreek}
\usepackage{slashed}
\usepackage{lineno}

\hypersetup{colorlinks=true, urlcolor=blue, citecolor=blue, filecolor=blue, linkcolor=blue}
\newcommand{\aZ}{\alpha Z}
\newcommand{\aZs}{(\alpha Z)^2}

\begin{document}

\title{Access to the kaon radius with kaonic atoms}

\date{\today}

\author{Niklas Michel}
\affiliation{Max-Planck-Institut f\"{u}r Kernphysik, Saupfercheckweg 1, 69117 Heidelberg, Germany}
\author{Natalia~S.~Oreshkina}
\email[Email: ]{Natalia.Oreshkina@mpi-hd.mpg.de} 
\affiliation{Max-Planck-Institut f\"{u}r Kernphysik, Saupfercheckweg 1, 69117 Heidelberg, Germany}

\begin{abstract}
We put forward a method for determination of the kaon radius from the spectra of kaonic atoms. 
We analyze the few lowest transitions and their sensitivity to the size of the kaon for ions in the nuclear charge range $Z=1-100$, taking into account finite-nuclear-size, finite-kaon-size, recoil and leading-order quantum-electrodynamic effects.
Additionally, the opportunities  of extracting the kaon mass and nuclear radii are demonstrated by examining the sensitivity of the transition energies in kaonic atoms.
\end{abstract}
% make the last sentense stromngeer

\maketitle

%%%%%%%%%%%%%%%%%%%%%%%%%%%%%%%%%%%%%%%%%%%%%%%%%%%%%%%%%%%%%%%%%%%%%%%%%%%%%%%%%%%%%%%%%%%%%%%%%%%%%%%%%%%%%%%%%%%%%%%%%%%%%%%%%%%%%%%%%%%%%%%%%%%%%%%%%%%%%%%%%%%%%%%%%%%%%%

{\it Introduction.}
While kaons are no elementary particles,  they represent themselves a very promising system to study, because they are the lightest meson with non-zero strangeness.
Despite their short lifetime of the order of $10^{-8}$~s, negatively charged kaons can nevertheless be captured by a nucleus and in this way  form a so-called kaonic atom.
Experimental studies of kaonic hydrogen and kaonic helium were performed lately by the DA$\Phi$NE  collaboration, testing kaon-nucleon strong interaction \cite{Bazzi2009,Bazzi2011,Bazzi2011_2,Bazzi2012}.
The corresponding theory lies at the interface between atomic, nuclear and particle physics, and was presented, e.g., by accurate QED prediction for the energies of the circular transitions for several kaonic atoms reported in Ref.~\cite{Santos2005}, by the strong contribution estimated in Ref.~\cite{Cheng1975,Gal2007}, and by the analysis of scattering amplitudes of kaonic atoms in Ref.~\cite{Friedman2013}.

It is generally  known, that experimental measurements in combination with  theoretical predictions  can unveil  yet unknown physical properties or constants, or improve those which are extremely important to know with high precision for further fundamental research.
However,  kaon mass and kaon radius values have not been updated since they were reported more than thirty years ago.
For kaon mass determination, exotic-atom x-ray spectroscopy has been used~\cite{Cheng1975,Gall1988}, whereas for the kaon radius has been measured by the direct scattering of kaons on electrons~\cite{Amendolia1986}.
Another exotic systems, namely muonic atoms, have been proved to be extremely sensitive to nuclear parameters, and therefore their studies allowed to retrieve information about atomic nuclei (see, e.g., Refs.~\cite{Pohl2010,Antognini2020}).
Being twice as heavy as muons, kaons in kaonic atoms feature even stronger dependence on nuclear parameters, making possible their extraction from spectroscopic data on exotic kaonic atoms.

In the current manuscript,
%Letter, 
we consider the spontaneous decay spectra of kaonic atoms for nuclear charges in the range $Z=1-100$ in order to establish how these systems can be used for the determination of the kaonic mass, kaonic radius and nuclear radius. 
This opens   access to the fundamental properties of kaons and giving a new path to probe essential properties of atomic nuclei.

%%%%%%%%%%%%%%%%%%%%%%%%%%%%%%%%%%%%%%%%%%%%%%%%%%%%%%%%%%%%%%%%%%%%%%%%%%%%%%%%%%%%%%%%%%%%%%%%%%%%%%%%%%%%%%%%%%%%%%%%%%%%%%%%%%%%%%%%%%%%%%%%%%%%%%%%%%%%%%%%%%%%%%%%%%%%%%

{\it Klein-Gordon equation.}
As a spinless particle, a kaon with a mass $m_K$ is described by the stationary Klein-Gordon equation (in the natural system of units, $\hbar = c = 1$) as \cite{Greiner}:

\begin{eqnarray}
\left[(E-V(\mathbf{r}))^2 + \Delta - m_K^2 \right] \varphi(\mathbf{r})=0.
\end{eqnarray}
In the case of a spherically-symmetric potential $V(\mathbf{r})=V(r)$, the angular variables can be separated from the radial ones as $\varphi(\mathbf{r})=R_l(r) / r \times {Y}_{lm}(\vartheta,\varphi)$, where $l$ and $m$ are the orbital quantum number and its projection, respectively. 
The angular part $Y_{lm}$ of the bound kaon wave function  consists of  spherical harmonics and therefore exactly coincides with that of a non-relativistic electron, described by the Sch\"odinger equation.
The radial part, represented as $R_l(r)=\phi_l(r)+\chi_l(r)$, also satisfies the Schr\"odinger-form set of equations:

\begin{subequations}\label{eq:KG_radial}
\begin{align}
\left[-\frac{D_l}{2m_K} +m_K+V(r)\right]\phi_l(r) + \frac{D_l}{2m_K} \chi_l(r) = E \phi_l(r),\\
\frac{D_l}{2m_K}\phi_l(r) + \left[\frac{D_l}{2m_K} -m_K+V(r)\right]\chi_l(r) = E \chi_l(r).
\end{align}
\end{subequations}
The differential operator $D_l$ acts as:
\begin{eqnarray*}
D_l(r) = \partial_r^2 -\frac{l(l+1)}{r^2}.
\end{eqnarray*}

Assuming the nucleus to be point-like and infinitely heavy, one can describe kaon-nucleus interaction with  the Coulomb potential $V = -\aZ/r$, where $\alpha = e^2/(4\pi)\approx 1/137$ is the fine-structure constant. 
Then, Eq.~\eqref{eq:KG_radial} can be solved analytically, resulting in the energies:
\begin{eqnarray}\label{eq:KG0_en}
\epsilon_{nl}(\aZ) = m_K\left( 1 + \cfrac{\aZs}{n_K-1/2+\mu} \right)^{-1/2}, 
\end{eqnarray}
where
\begin{eqnarray*}
n_K &=& n-l, \\
\mu &=& \sqrt{(l+1/2)^2 - \aZs},
\end{eqnarray*}
and the wavefunctions:
\begin{eqnarray}\label{eq:KG0_wf}
\psi_{nl}(r) = N_0\rho^{\mu+1/2} e^{-\rho/2}F_1(1-n_K, 2\mu+1, \rho),
\end{eqnarray}
where 
\begin{eqnarray*}
\rho &=& 2 r \sqrt{1-\epsilon^2/m_K^2},\\
\end{eqnarray*}
and $N_0$ is the normalization constant:
\begin{eqnarray*}
N_0 &=& \frac{(1-\epsilon^2/m_K^2)^{1/4}}{\Gamma(2\mu+1)} \\ &\times &\left(\frac{2\Gamma(n_K-1+2\mu+1)} {(n_K-1)! [2(n_K-1)+2\mu +1 ]}\right)^{1/2}.
\end{eqnarray*}
Eq.~\eqref{eq:KG0_en} contains a singularity in the denominator, and for $l=0$ it  breaks at $\aZ=1/2$, or at $Z\approx 69$. 
This indicates that the point-like-nucleus approximation is not valid anymore, and  one has to consider a more realistic nuclear model, including finite nuclear size effects.

%%%%%%%%%%%%%%%%%%%%%%%%%%%%%%%%%%%%%%%%%%%%%%%%%%%%%%%%%%%%%%%%%%%%%%%%%%%%%%%%%%%%%%%%%%%%%%%%%%%%%%%%%

{\it Finite-nuclear-size effect.}
One of the simplest nuclear models is a homogeneously charged sphere, with the corresponding charge density of the nucleus
\begin{equation}
\rho(r) = \frac{3Ze}{4\pi r_0^3}\theta(r_0-r).
\end{equation} 
Here $Z$ is the nuclear charge   and $r_0$ is the effective radius of the nucleus, associated  with a root-mean-square~(RMS) radius of the nucleus as 
\begin{equation}
r_0 = \sqrt{\frac{5}{3}\langle r^2\rangle}. \label{radius of sphere}
\end{equation} 
The interaction between electron and nucleus can be therefore described by the potential
\begin{equation}\label{eq:sphere}
V_{\rm sphere}(r)=
\begin{cases}
 -\cfrac{Z\alpha}{2r_0}\biggl(3-\cfrac{r^2}{r_0^2}\biggr), & \textrm{while } r \leq r_0, %\textrm{ [Region \rom{1}}]; 
 \\
 -\cfrac{Z\alpha}{r}, & \textrm{while } r>r_0. % \textrm{ [Region \rom{2}}].
\end{cases}
\end{equation}
With this potential, the Eq.~\eqref{eq:KG_radial} can be in principle solved semi-analytically in analogy to Ref.~ \cite{Shabaev1984,Patoary}   for electrons and muons, described by the Dirac equation.
However, even for a muon, which is more than two times lighter than a kaon, and therefore is located on a larger distance from the nucleus, the semi-analytical method in a first order of FNS correction turns out to be not sufficient~\cite{Patoary,Michel2017}. 

%%%%%%%%%%%%%%%%%%%%%%%%%%%%%%%%%%%%%%%%%%%%%%%%%%%%%%%%%%%%%%%%%%%%%%%%%%%%%%%%%%%%%%%%%%%%%%%%%%%%%%%%%

{\it Finite-kaon-size effect.}
Additionally to the finite-nuclear-size effect, one can take into account the finite-kaon-size (FKS) effect.
To estimate the order of magnitude of FKS, we used a comparably simple two-sphere approach to build a potential, presented in Ref.~\cite{Mitra}.

Denoting the radius of the nucleus as $R_N$, and the radius of the kaon as $R_K$, we assume that these two spheres interact without deformation via electromagnetic forces.
Then, {denoting the radii ratio as $\lambda = R_K/R_N$,} three different regions for $\rho = r/R_N$ should be considered: 
\begin{itemize}
\item[(i)]$\qquad 0 \leq \rho \leq 1-\lambda$, the kaon is totally inside the nucleus,
\item[(ii)] $\quad 1-\lambda \leq \rho \leq 1+\lambda$, the kaon and the nucleus partly overlap, and
\item[(iii)] $\qquad \rho \geq 1+\lambda$, the kaon is totally outside the nucleus.
\end{itemize}
The corresponding potential $V(\rho)$ is determined as:
\begin{align}\label{eq:two_spheres}
V_{2\rm spheres}(r)=
\begin{cases}
 -V_0(C_0-\rho^2), &\textrm{(i),}  \\
 -\cfrac{3}{8}\cfrac{V_0}{\lambda^3}  
   \biggl(\cfrac{C_1}{\rho} + \sum\limits_{k=0}^5 C_{k+2}\rho^k\biggr), 
   &\textrm{(ii),} \\
- V_0/\rho,  &\textrm{(iii).}  \\
\end{cases}
\end{align}
Here $V_0 = -\aZ/R_N$, and
the coefficients in Eq.~\eqref{eq:two_spheres} are determined as:
\begin{align*}
C_0 &= 3/2 - 3\lambda^2/10,\\
C_1 &= (1-9\lambda^2+16\lambda^3-9\lambda^4)/12,\\
C_2 &= (-2+10\lambda^2+10\lambda^3-2\lambda^5)/5,\\
C_3 &= (3+6\lambda^2 +3\lambda^4)/4,\\
C_4 &= (-2-9\lambda^2 -2\lambda^3)/3,\\
C_5 &= (1+\lambda^2)/4, \\
C_6 &= 0, \\
C_7 &= -1/60.
\end{align*}
By calculating energies of a given state with a homogeneously-charged sphere~\eqref{eq:sphere} or two-spheres~\eqref{eq:two_spheres} potential, one can evaluate the FKS effect:
\begin{equation}\label{eq:fks}
\delta_{\rm FKS} = 1 - \frac{\epsilon_{nl}[V_{2\rm spheres}]}{\epsilon_{nl}[V_{\rm sphere}]}. 
\end{equation} 

%%%%%%%%%%%%%%%%%%%%%%%%%%%%%%%%%%%%%%%%%%%%%%%%%%%%%%%%%%%%%%%%%%%%%%%%%%%%%%%%%%%%%%%%%%%%%%%%%%%%%%%%%

{\it Quantum-electrodynamic effects.}
Another important contribution to the energies of kaonic atoms originates from the quantum-electrodynamics (QED) corrections.
In the first order in $\alpha$, there are self-energy (SE) and vacuum polarization (VP) corrections.
For hydrogen-like electronic ions these two corrections are of the same order of magnitude.
However, even for muonic atoms it is already not so: due to the large muon-electron mass ratio, the VP with a virtual electron-positron pair is a few orders of magnitude larger than the VP with a virtual muon-antimuon pair or SE correction (see, e.g.,~\cite{Borie,Michel2017}).
The same stands also for kaonic atoms, and therefore the leading QED correction can be described by the Uehling potential \cite{Elizarov2005}:
\begin{align}
V_{\text{Uehl}}(r)=-\alpha \frac{2\alpha}{3\pi}\int_0^\infty \text{d}r^{\prime}\,4\pi \rho(r^\prime)\int_1^\infty \text{d}t\,\left( 1+\frac{1}{2t^2} \right)\nonumber\\
\times\frac{\sqrt{t^2-1}}{t^2} \frac{\text{exp}(-2m_e|r-r^\prime|t)-\text{exp}(-2m_e(r+r^\prime)t)}{4m_er t},
\label{eq:uehl}
\end{align}
with $m_e$ being the mass of an electron.
%
%This potential can be also included in Klein-Gordon equation, and calculated energies would therefore account  also leading QED effect at all orders.
%\textcolor{green!70!black}
{
In our calculation, this potential has also been included in the Klein-Gordon equation. 
Therefore, the calculated energies account for the leading QED effect to all orders.
}

%%%%%%%%%%%%%%%%%%%%%%%%%%%%%%%%%%%%%%%%%%%%%%%%%%%%%%%%%%%%%%%%%%%%%%%%%%%%%%%%%%%%%%%%%%%%%%%%%%%%%%%%%

{\it Recoil.} 
Due to the large mass of a kaon compared to a proton, the recoil effect is also extremely important for very light ions, but becomes negligible for middle and heavy ions.
To evaluate it, we used a simple reduced-mass formalism \cite{Landau}, replacing the nuclear mass $m_N$ with
\begin{equation}
m_r = \frac{m_Nm_K}{m_N+m_K}.
\end{equation}
Standard atomic weights \cite{Atomic2013} have been used in the current manuscript.

%%%%%%%%%%%%%%%%%%%%%%%%%%%%%%%%%%%%%%%%%%%%%%%%%%%%%%%%%%%%%%%%%%%%%%%%%%%%%%%%%%%%%%%%%%%%%%%%%%%%%%%%%

{\it Sensitivities.}
Taking into account FNS, FKS, the leading QED and recoil effects,  we calculated a kaonic atom spectrum. 
To characterize how  values of physical observable change depending on the parameters of the theory used, one can introduce sensitivity coefficients:
\begin{eqnarray}
\frac{\Delta E}{E} &= K_R \cfrac{\Delta R_N}{R_N} + K_m \cfrac{\Delta m_K}{m_K}.
\end{eqnarray}
By varying different parameters of our calculations, we can estimate the corresponding sensitivity factors, similarly as it was done in earlier works for other physical constants, e.g. in Ref.~\cite{Oreshkina2017}.
We use only the most general sensitivity coefficients $K_R$ and $K_m$ in our current work, since the kaonic atoms spectra feature complicated non-linear dependence on %$R_K$,
 $R_N$ and $m_K$.
For example, the mass of the kaon is a scaling factor for all energies, however, it is also should be included in the nuclear potential via scaling of the radius and reduced mass.
Analogously with the nuclear radius: for simple atomic systems, like electronic H-like ions, one can describe FNS effect via simple term $\Delta E_{\rm FNS} \propto (\alpha Z)^2$ \cite{Shabaev_1993}.
For kaonic atoms the FNS correction has much higher impact, and therefore one should take into account also higher order terms (see, e.g. \cite{Karshenboim2018,Michel2019}).
As an outcome, the sensitivity coefficients are rather ion- and transition- dependent and  can vary in a sizable range.

%%%%%%%%%%%%%%%%%%%%%%%%%%%%%%%%%%%%%%%%%%%%%%%%%%%%%%%%%%%%%%%%%%%%%%%%%%%%%%%%%%%%%%%%%%%%%%%%%%%%%%%%%

%\textcolor{green!70!black}
{
{\it Strong shift.} The strong interaction effects in the spectra of kaonic atoms are also extremely important and can significantly change the binding and transition energies.
Thus, for kaonic lead Pb${^{82}}$, the binding energy of the $1s$ state is $E_{1s}[\text{Pb}^{82}]=-17$~MeV.
The hadronic contribution of $\sim 10$~MeV desreaces it to only $E_{1s}[\text{Pb}^{82}]=-7$~MeV~\cite{Cheng1975, Santos2005}. 
Analogously, the strong shift would decrease transition energies for all other kaonic atoms.
In the following, we will focuse only on QED calculations, however, in the future, it should be indisputably taken into account.
}

%%%%%%%%%%%%%%%%%%%%%%%%%%%%%%%%%%%%%%%%%%%%%%%%%%%%%%%%%%%%%%%%%%%%%%%%%%%%%%%%%%%%%%%%%%%%%%%%%%%%%%%%%%%%%%%%%%%%%%%%%%%%%%%%%%%%%%%%%%%%%%%%%%%%%%%%%%%%%%%%%%%%%%%%%%%%%%

\begin{figure} [t]
\begin{center}
\includegraphics[width=0.99\columnwidth]{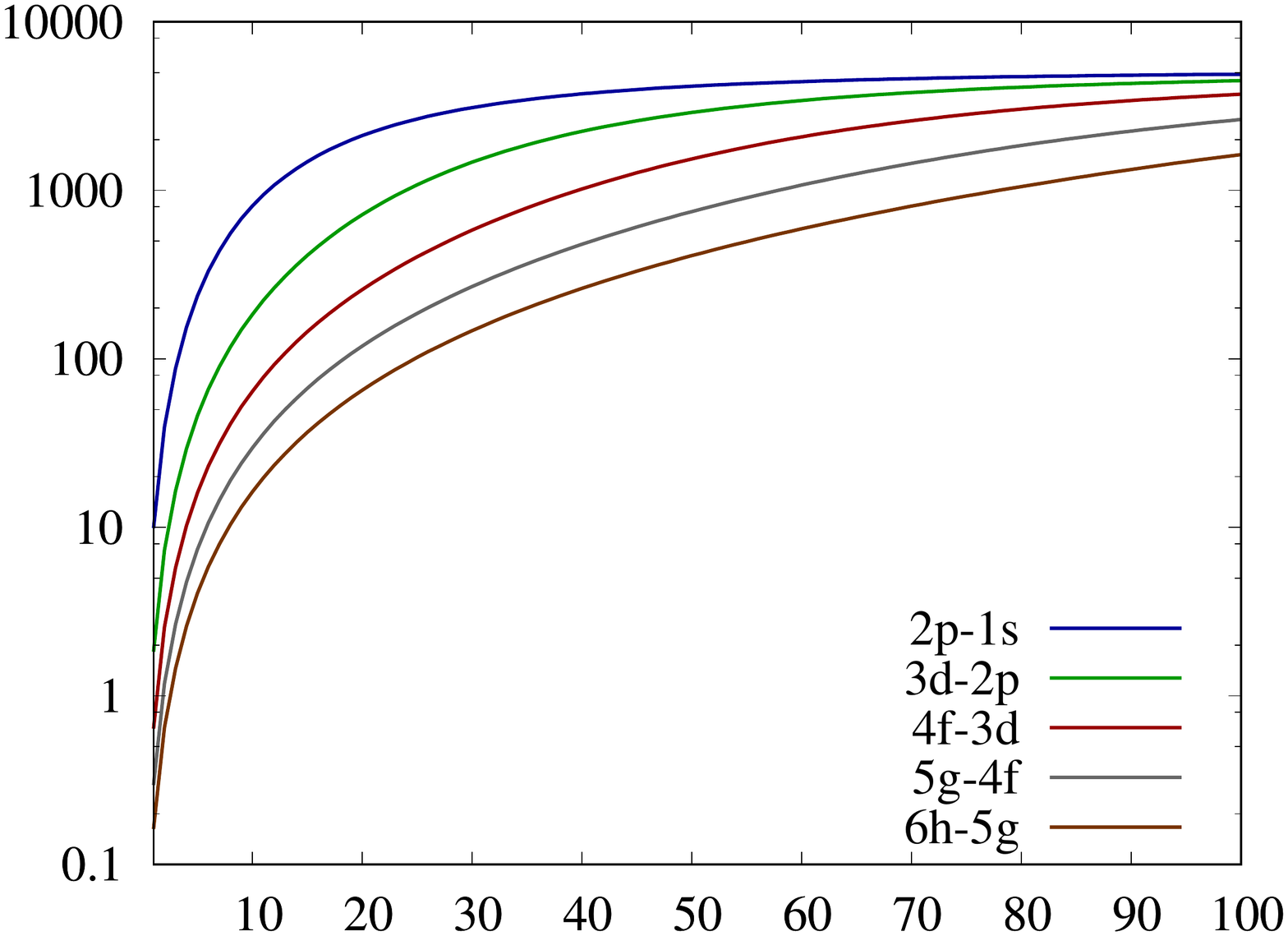} 
\caption {Transition energies of the first few circular transitions for kaonic atoms in the range $Z=1-100$, in keV. }\label{fig:energies}
\end{center}
\end{figure}

{\it Results. }
Using the above described method, we calculated the spectra and transition energies for $Z=1-100$. 
In order to work with the most general expressions, we assumed the nuclear radius  to be $R_{N} = 1.2Z^{1/3}$, and for the mass of the nucleus the standard atomic weight has been used. 
Such simple assumptions allowed us to analyze  the general trends for our observables, however, for high precision calculation aiming the access to  nuclear or particle parameters from an experiment, one has to use tabulated nuclear data, e.g.~\cite{Angeli} for root-mean-square radius.
The transition energies, including of FNS, FKS, the leading QED and recoil effect for the circular transitions from $2p \rightarrow 1s$ up to  $6h \rightarrow 5g$, are plotted  as functions of nuclear charge $Z=1-100$ in Fig.~\ref{fig:energies}. 
In Table \ref{tab:sens}, the same energies, FKS correction $\delta_{\rm FKS}$ and the   sensitivities to  the nuclear radius $K_R$ and to the mass of a kaon $K_m$ are listed for few kaonic atoms: helium He$^{2}$, titanium Ti$^{22}$, xenon Xe$^{54}$ and uranium U$^{92}$. 
Our value for $3d\rightarrow 2p$ transition  of 6.465~keV is in a perfect agreement with previously reported experimental and theoretical values~\cite{Bazzi2009}.

\begin{table}
\begin{center} % r@{}l
\begin{tabular} {l r r@{}l r@{}l r@{}l r@{}l}
\hline \hline
Ion & Transition	&\qquad $\Delta E$, \, &[keV] &  $\delta_{\rm FKS}$&, [\%]&$K$&$_R$ 	& $K$&$_m$ \\
\hline
He$^2$	& $2p \rightarrow 1s$	& 34.&84	& 0&.07		& -0&.01	& 0.&99 \\
		& $3d \rightarrow 2p$	& 6.&465	& 2&[-5] 	& -2&[-6] 	& 1.&00 \\
		& $4f \rightarrow 3d$	& 2.&259	& 0& 		& 0&		& 1.&00 \\
		& $5g \rightarrow 4f$	& 1.&045	& 0& 		& 0&		& 1.&00 \\
		& $6h \rightarrow 5g$	& 0.&5734	& 0&		& 0&		& 1.&00 \\
\hline 
Ti$^{22}$	& $2p \rightarrow 1s$	& 2315.&	& 0&.80	 & -0&.71	& 0.&27	 \\
			& $3d \rightarrow 2p$	& 846.&8 	& 0&.20  & -0&.12	& 0.&87	 \\
			& $4f \rightarrow 3d$	& 308.&6	& 5&[-3] & -3&[-3]	& 0.&99	 \\
			& $5g \rightarrow 4f$	& 142.&5	& 2&[-5] & -9&[-6]	& 1.&00	 \\
			& $6h \rightarrow 5g$	& 78.&06	& 3&[-8] & 0& 		& 1.&00	 \\
\hline 
Xe$^{54}$	& $2p \rightarrow 1s$	& 4253.&	& 0&.49	 & -1&.2	& -0.&23 \\
			& $3d \rightarrow 2p$	& 3109.&	& 0&.57  & -0&.80	& 0.&18 \\
			& $4f \rightarrow 3d$	& 1746.&	& 0&.26  & -0&.25	& 0.&74 \\
			& $5g \rightarrow 4f$	& 868.&8	& 0&.03  & -0&.02 	& 0.&98 \\
			& $6h \rightarrow 5g$	& 476.&3	& 6&[-4] & -4&[-4] 	& 1.&00 \\
\hline 
U$^{92}$	& $2p \rightarrow 1s$	& 4825.&	& 0&.23 & -1&.4		& -0.&39 \\
			& $3d \rightarrow 2p$	& 4340.&	& 0&.41 & -1&.2		& -0.&21 \\
			& $4f \rightarrow 3d$	& 3470.&	& 0&.46 & -0&.84 	&  0.&15 \\
			& $5g \rightarrow 4f$	& 2322.&	& 0&.26 & -0&.34 	&  0.&65 \\
			& $6h \rightarrow 5g$	& 1386.&	& 0&.05 & -0&.05	& 0.&95 \\
\hline 
\hline 
\end{tabular} 
\caption{Transition energies $\Delta E$, finite-kaon-size effect $\delta_{\rm FKS}$, and sensitivities to the nuclear radius $K_R$ and mass of a kaon for $K_m$ for the first few circular transitions in kaonic helium He$^2$, titanium Ti$^{22}$, xenon Xe$^{54}$,  and  uranium U$^{92}$. 
The number in square parenthesis indicates the power of 10. } 
\label{tab:sens}
\end{center}
\end{table}

%%%%%%%%%%%%%%%%%%%%%%%%%%%%%%%%%%%%%%%%%%%%%%%%%%%%%%%%%%%%%%%%%%%%%%%%%%%%%%%%%%%%%%%%%%%%%%%%%%%%%%%%%%%%%%%%%%%%%%%%%%%%%%%%%%%%%%%%%%%%%%%%%%%%%%%%%%%%%%%%%%%%%%%%%%%%%%

\begin{figure} [t]
\begin{center}
\includegraphics[width=0.99\columnwidth]{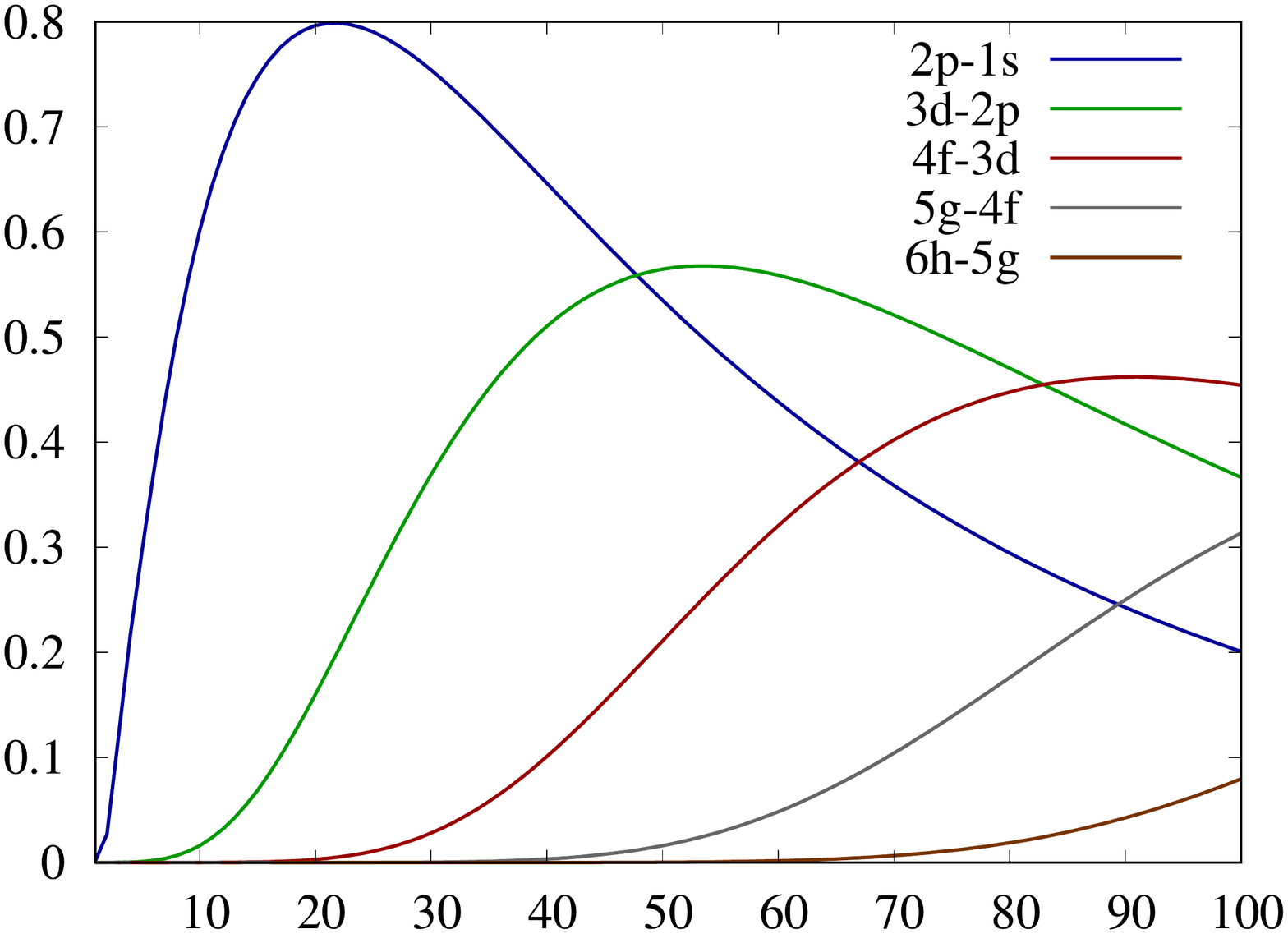} 
\caption {Finite kaon size effect $\delta_{\rm FKS}$ (in percent), estimated within two-spheres model, for the transition energies of the first few circular transitions for kaonic atoms in the range $Z=1-100$.}
\label{fig:fks}
\end{center}
\end{figure}

{\it Determination of kaon's radius.}
In Fig. \ref{fig:fks} one can see the relative FKS correction~(in percent) to the transition energy as a function of nuclear charge $Z$. 
Since the potential in Eq.~\eqref{eq:two_spheres} is a function of $R_K/R_N$, and the radius of the kaon $R_K = 0.40$ fm is smaller than any nuclear radius,
one could naively expect, that the FKS effect would be the largest when the kaon-nucleus size ratio is minimal, e.g. for hydrogen.
However, as one can see from the Fig.~\ref{fig:fks}, this is not the case.
The relative FKS effect to the $2p\rightarrow 1s$ transition grows with $Z$, reaching the maximum value of 0.8\% at $Z\approx 22$~\footnote{within our simple model. With more sophisticated calculations, the position of the maximum can slightly shift, not changing nevertheless the general trend.}, and then it starts to decrease.
Also, unlike  the FNS effect, which is always maximized for the $1s$ shell, and getting smaller and finally simply negligible for the higher atomic shells, we can observe quite a different trend for the FKS effect.  
All other transitions exhibit the same qualitative behavior with respect to the FKS effect as the $2p\rightarrow 1s$ transition, however with different positions and values of their maxima.
Thus, for $3d\rightarrow 2p$ transition the FKS effect has its maximum of 0.57\% at $Z\approx 54$, and $4f\rightarrow 3d$ the maximum of 0.46\% can be reached at $Z\approx 92$.
%\textcolor{green!70!black}
{
Since the strong shift decreases the transition energies, accounting previously neglected strong contributions would lead to the further enhancement in the transitions' sensitivity to the FKS effect. 
}
Therefore, it opens various possibilities for the determination of the size of the kaon based on the different transitions of the kaonic atoms with a different nuclear charge $Z$.  

%%%%%%%%%%%%%%%%%%%%%%%%%%%%%%%%%%%%%%%%%%%%%%%%%%%%%%%%%%%%%%%%%%%%%%%%%%%%%%%%%%%%%%%%%%%%%%%%%%%%%%%%%%%%%%%%%%%%%%%%%%%%%%%%%%%%%%%%%%%%%%%%%%%%%%%%%%%%%%%%%%%%%%%%%%%%%%

\begin{figure} [t]
\begin{center}
\includegraphics[width=0.99\columnwidth]{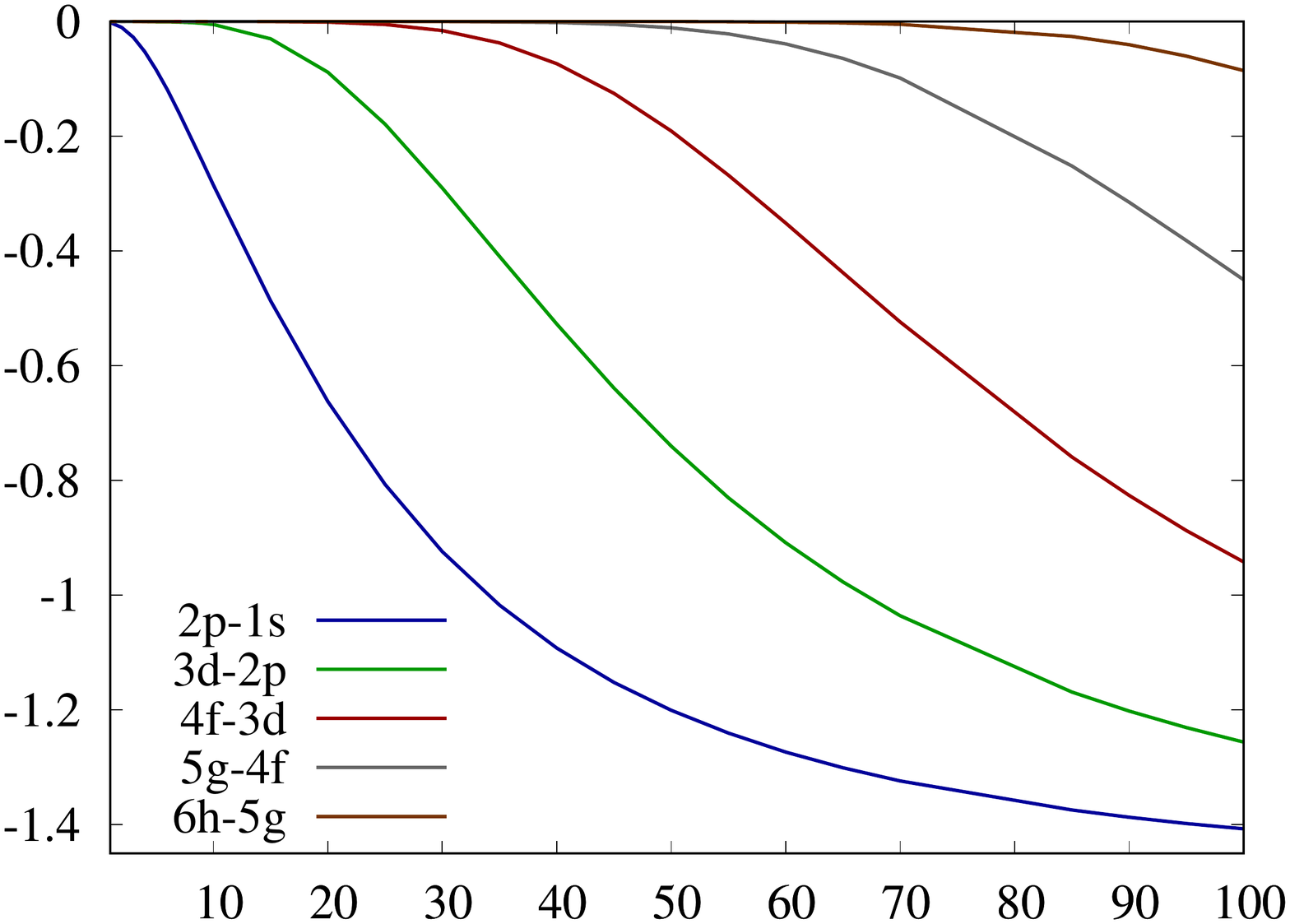} 
\caption {Sensitivity coefficient $K_R$  to the radius of the nucleus, for the transition energies of the first few circular transitions for kaonic atoms in the range $Z=1-100$.}
\label{fig:sens_r}
\end{center}
\end{figure}

{\it Extraction of nuclear radii.}
As one can see from Table~\ref{tab:sens} and from Fig.~\ref{fig:sens_r}, for all ions the transitions to the low-lying states are sensitive to the nuclear radius, and therefore can be used for its extraction.
However, since in the ground state the energy of a kaon is also affected by other  interactions with the nucleus, which are not so easy to quantify (strong and weak interactions, nuclear-polarization correction, {\it etc.}), the transition $2p \rightarrow 1s$ can be not very suitable for this procedure.
Therefore, the next transition $3d \rightarrow 2p$, and for heavy ions even $4f \rightarrow 3d$, can provide the necessary information about nuclear radius.

%%%%%%%%%%%%%%%%%%%%%%%%%%%%%%%%%%%%%%%%%%%%%%%%%%%%%%%%%%%%%%%%%%%%%%%%%%%%%%%%%%%%%%%%%%%%%%%%%%%%%%%%%%%%%%%%%%%%%%%%%%%%%%%%%%%%%%%%%%%%%%%%%%%%%%%%%%%%%%%%%%%%%%%%%%%%%%

\begin{figure} [t]
\begin{center}
\includegraphics[width=0.99\columnwidth]{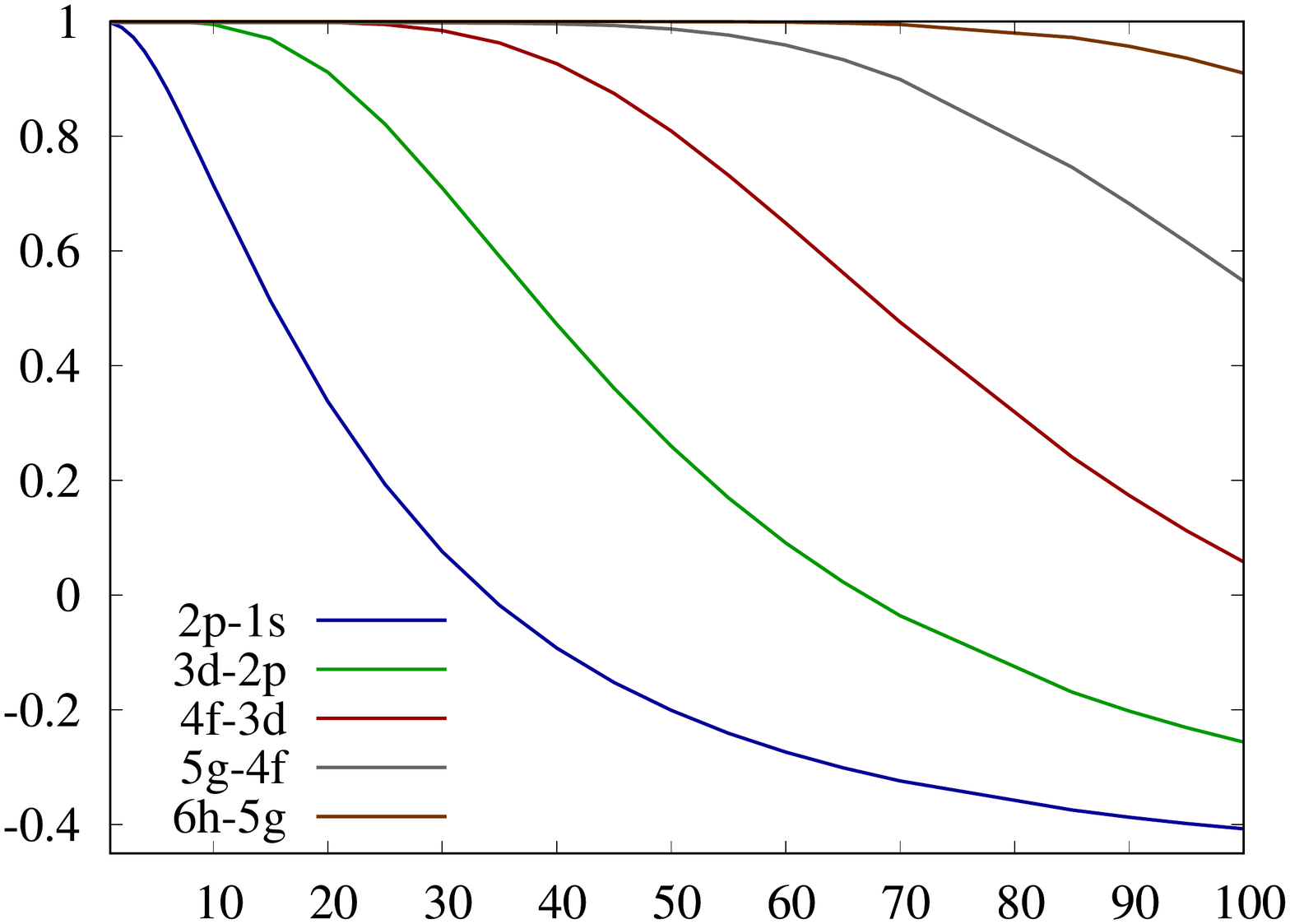} 
\caption {Sensitivity coefficient $K_m$  to the kaon mass, for the transition energies of the first few circular transitions for kaonic atoms in the range $Z=1-100$.}
\label{fig:sens_m}
\end{center}
\end{figure}

{\it Extraction of the mass of a kaon.}
Due to the complicated dependence of the energy on kaons mass via the nuclear radius, the sensitivity coefficient $K_m$ differs from unity, especially for the lowest-lying transitions, see Fig.~\ref{fig:sens_m}. 
However, even for heaviest element considered, it is close to unity for the transitions starting with $6h \rightarrow 5g$, and therefore the analysis of kaonic atom spectra can be used for the determination of its mass.
The fact that the dependence holds for any nuclei can be used to enlarge the statistics and choose the system with the most suitable parameters for an experiment.
A similar procedure was already used  before in Ref.~\cite{Gall1988}, however, with continuous progress in both experimental technique and theoretical calculations one can improve the existing accuracy.

%%%%%%%%%%%%%%%%%%%%%%%%%%%%%%%%%%%%%%%%%%%%%%%%%%%%%%%%%%%%%%%%%%%%%%%%%%%%%%%%%%%%%%%

{\it Further improvements.}
So far, we only showed the principal  idea of  using kaonic atoms for the extraction of  particle and nuclear  parameters, considering the leading size and QED effects.
However, for high-precision theoretical predictions to be compared with experimental data, one has to take into account effects already calculated in this manuscript and other effects, with higher accuracy, similarly as it was done e.~g. for muonic atoms in Ref.~\cite{Antognini2020}.
First of all, 
%\textcolor{green!70!black}
{
the strong and weak interaction contributions can be either taken from the previously reported data~\cite{Cheng1975,Gal2007} or calculated to the required accuracy with current state-of-the-art methods.
Then, }
the FNS effect can be calculated with Fermi or even deformed Fermi nuclear potential~\cite{Michel2017}, or with more realistic predictions based on the Skyrme-type nuclear potential~\cite{Valuev2020}, not forgetting about nuclear deformation correction~\cite{Michel2019nd}. 
There is a room for an improvement in the evaluation of FKS as well, from a simple two-sphere model to a more sophisticated and realistic one.
The higher-order QED effects, such as self-energy, Wichmann-Kroll, K\"{a}ll\'{e}n-Sabry, muonic and hadronic Uehling potentials~\cite{Santos2005, Michel2019nd, Dizer} should be also included. 
The electron screening effects were shown to be negligible in the spectra of muonic atoms~\cite{Vogel_1973,Michel2017}, therefore we expect them to be even smaller for kaonic atoms, which has to be nevertheless checked.
The recoil effect should be included within a rigorous relativistic approach~\cite{Santos2005, Recoil_1983}, for more precise values for  light kaonic atoms.
Finally, the effects of nuclear polarization have to be taken into account, as it was done, e.g., in Refs.~\cite{Nefiodov1996,HalilReduced}. %, whereas the strong and weak interaction contributions can be taken from Ref.~\cite{Gal2007}.
%\textcolor{green!70!black}
{
For all above mentioned atomic structure effects, the {1\textperthousand} or better precision can be reached, which makes our suggestions quite realistic. 
}

%%%%%%%%%%%%%%%%%%%%%%%%%%%%%%%%%%%%%%%%%%%%%%%%%%%%%%%%%%%%%%%%%%%%%%%%%%%%%%%%%%%%%%%

{\it Conclusions.}
We considered  kaonic atoms with the nuclear charge $Z=1-100$.
Taking into account finite-nuclear size, finite-kaon size, leading-order quantum-electrodynamic and recoil effects, we calculated transition energies and sensitivity coefficients to the nuclear radius and mass of a kaon.
We analyzed the finite-kaon-size effect, showing that the value of the kaon's radius can be extracted with almost equal efficiency from few different transitions and for few different ions.
Similarly, the decay spectra of kaonic atoms can be used for the determination of nuclear radii and the mass of the kaon.
%\textcolor{green!70!black}
{
The choice of the most suitable system for any of these purposes should be made based on many important parameters, such as the accuracy of theoretical predictions, the natural linewidths, experimental accessibility of the particular nuclei, and the possibility to carry high-precision measurements.
}
Concluding, the knowledge of atomic structure of kaonic atoms would give access to the fundamental nuclear and particle parameters, and therefore could motivate new experiments and high-precision calculations.

{\it Acknowledgements.}
N.~M. acknowledges  support  by  the IMPRS.
The Authors thank Catalina Oana Curceanu for drawing our attention to  kaonic atoms, and Vincent Debierre and Zolt{\'a}n Harman for discussion and comments.  

\bibliography{refs_kaon}

\end{document}